\documentclass[journal]{IEEEtran}
\usepackage{cite}
\usepackage[squaren, Gray, cdot]{SIunits}
\usepackage[usenames,dvipsnames,svgnames,table]{xcolor}
\usepackage{hyperref}
\usepackage{algorithm}
\usepackage{algpseudocode}
\usepackage{placeins}

\ifCLASSINFOpdf
   \usepackage[pdftex]{graphicx}
\else
\fi
\usepackage{amsmath}

\usepackage{array}

\begin{document}
\title{
Implementation  of  Ternary Weights  \\with  Resistive RAM Using a Single \\Sense Operation per Synapse
\author{
Axel Laborieux,~\IEEEmembership{Student Member,~IEEE,}
Marc Bocquet, 
Tifenn Hirtzlin,~\IEEEmembership{Student Member,~IEEE,} \\
Jacques-Olivier Klein,~\IEEEmembership{Member,~IEEE,} 
Etienne Nowak, 
Elisa Vianello,~\IEEEmembership{Member,~IEEE,}\\  
Jean-Michel Portal~\IEEEmembership{Member,~IEEE,} 
and Damien Querlioz~\IEEEmembership{Member,~IEEE} 
}
\thanks{Axel Laborieux, Tifenn Hirtzlin,  Jacques-Olivier Klein, and Damien Querlioz are with  Universit\'e Paris-Saclay, CNRS, Centre de Nanosciences et de Nanotechnologies, 91120 Palaiseau, France. email: damien.querlioz@c2n.upsaclay.fr}
\thanks{Marc Bocquet and Jean-Michel Portal are with Institut Mat\'eriaux Micro\'electronique Nanosciences de Provence, Univ. Aix-Marseille et Toulon, CNRS, France.}
\thanks{Etienne Nowak and Elisa Vianello are with Universit\'e Grenoble-Alpes, CEA, LETI, Grenoble, France.}
\thanks{This work was supported by the ERC Grant NANOINFER (715872) and  the ANR grant NEURONIC (ANR-18-CE24-0009).}
}

\IEEEoverridecommandlockouts
\maketitle
\begin{abstract}
The design of systems implementing low precision neural networks with emerging memories such as resistive random access memory (RRAM) is a significant lead for reducing the energy consumption of artificial intelligence.  To achieve maximum energy efficiency in such systems, logic and memory should be integrated as tightly as possible. In this work, we focus on the case of ternary neural networks, where synaptic weights assume ternary values. We propose a two-transistor/two-resistor memory architecture employing a precharge sense amplifier, where the weight value can be extracted in a single sense operation.  Based on experimental measurements on a hybrid 130~nm CMOS/RRAM chip featuring this sense amplifier, we show that this technique is particularly appropriate at low supply voltage, and that it is resilient to process, voltage, and temperature variations. We characterize the bit error rate in our scheme.  We  show based on neural network simulation on the CIFAR-10 image recognition task that the use of ternary neural networks significantly increases neural network performance, with regards to binary ones, which are often preferred for inference hardware. We finally evidence that the neural network is immune to the type of bit errors observed in our scheme, which can therefore be used without error correction.
\end{abstract}


\IEEEpeerreviewmaketitle

\section{Introduction}

Artificial Intelligence has made tremendous progress in recent years due to the development of deep neural networks. 
Its deployment at the edge, however, is currently limited by the high power consumption of the associated algorithms \cite{xu2018scaling}.
Low precision neural networks are currently emerging as a solution, as they allow the development of low power consumption hardware specialized in deep learning inference \cite{hubara2017quantized}.
The most extreme case of low precision neural networks, the Binarized Neural Network (BNN), also called XNOR-NET, is receiving particular attention as it is especially efficient for hardware implementation: both synaptic weights and neuronal activations assume only binary values \cite{courbariaux2016binarized,rastegari2016xnor}.
Remarkably, this type of neural network can achieve high accuracy on vision tasks \cite{lin2017towards}.
One particularly investigated lead is to fabricate hardware BNNs with emerging memories such as resistive RAM or memristors
\cite{bocquet2018,yu2016binary,giacomin2019robust,sun2018xnor,zhou2018new,natsui2018design,tang2017binary, lee2019adaptive}.
The low memory requirements of BNNs, as well as their reliance on simple arithmetic operations, make them indeed particularly adapted for ``in-memory'' or ``near-memory'' computing approaches, which achieve superior energy-efficiency by avoiding the von~Neumann bottleneck entirely.

Ternary neural networks \cite{alemdar2017ternary} (TNN, also called Gated XNOR-NET, or GXNOR-NET \cite{deng2018gxnor}), which add the value $0$ to synaptic weights and activations, 
are also considered for hardware implementations \cite{ando2017brein,prost2017scalable,li2017design,pan2019skyrmion}.  
They are comparatively receiving less attention than binarized neural networks, however. In this work, we highlight that implementing TNNs does not necessarily imply considerable overhead with regards to BNNs.
We introduce a two-transistor/two-resistor  memory architecture for TNN implementation. The array uses a precharge sense amplifier for reading weights, and the ternary weight value can be extracted in a single sense operation, by exploiting the fact that latency of the sense amplifier depends on the resistive states of the memory devices.
This work extends a  hardware developed for the energy-efficient 
implementation of BNNs \cite{bocquet2018}, where the synaptic weights are implemented in a differential fashion.
We, therefore, show that it can be extended to TNNs without overhead on the memory array.

The contribution of this work is as follows. After presenting the background of the work (section~\ref{sec:background}):
\begin{itemize}
\item We demonstrate experimentally, on a fabricated 130~nm RRAM/CMOS hybrid chip, a  strategy for implementing ternary weights using a precharge sense amplifier, which is particularly appropriate when the sense amplifier is operated at low supply voltage
(section~\ref{sec:circuit}).
\item We analyze  the bit errors of this scheme experimentally and their dependence on the RRAM programming conditions (section~\ref{sec:programmability}). 
\item We verify the robustness of the approach to process, voltage, and temperature variations (section~\ref{sec:PVT}).
\item We carry simulations that show the superiority of TNNs over BNNs on the canonical CIFAR-10 vision task, and evidence the error resilience of hardware TNNs (section~\ref{sec:network}).
\item We discuss the results, and compare our approach with the idea of storing three resistance levels per device.
\end{itemize}

Partial and preliminary results of this work have been presented at a conference \cite{laborieux2020low}. 
This journal version adds the experimental characterization of bit errors in our architecture, supported by a comprehensive analysis of the impact on process, voltage, and temperature variations, and their impact at the neural network level, together with a detailed analysis of the use of ternary networks over binarized ones.


\section{Background}
\label{sec:background}
 
 \begin{figure}[t]
	\centering
	\includegraphics[width=3.0in]{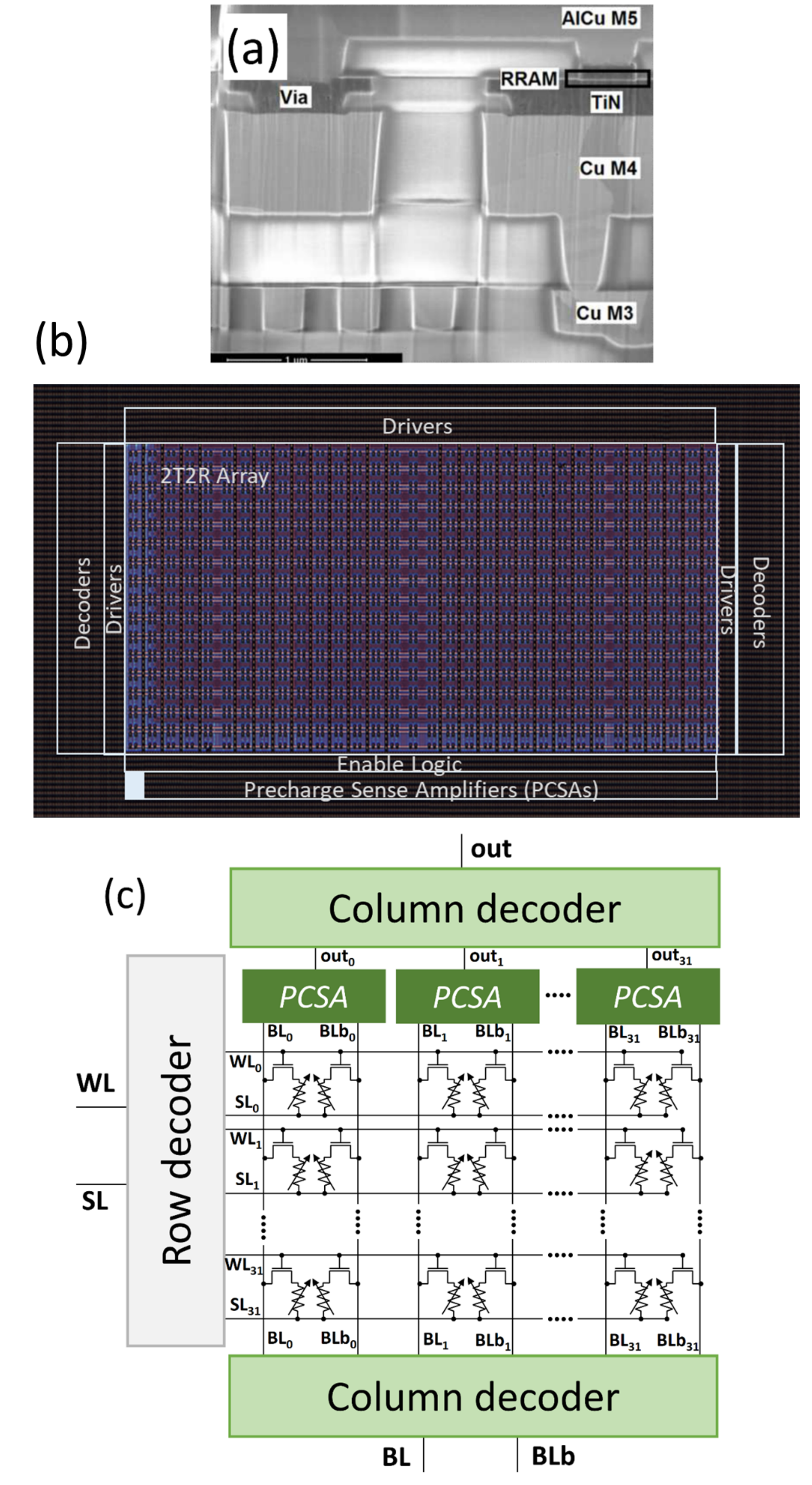}
	\caption{(a) Electron microscopy image of a hafnium oxide resistive memory cell (RRAM) integrated in the backend-of-line of a $130\nano\meter$ CMOS process. (b) Photograph and (c) simplified schematic of the hybrid CMOS/RRAM test chip characterized in this work.
	The white rectangle in (b) materializes a single PCSA.}
	\label{fig:testchip}
\end{figure}
 
The main equation in conventional neural networks is the computation of the neuronal activation $A_j =  f \left( \sum_i W_{ji}X_i \right),$ where $A_j$, the synaptic weights $W_{ji}$, and input neuronal activations $X_i$ assume real values, and $f$ is a non-linear activation function.
Binarized neural networks (BNNs) are a considerable simplification of conventional neural networks, in which all neuronal activations ($A_j$, $X_i$) and synaptic weights $W_{ji}$  can only take binary values meaning $+1$ and $-1$. 
Neuronal activation then becomes:
\begin{equation}
\label{eq:activ_BNN}
    A_j = \mathrm{sign} \left( \sum_i  XNOR \left( W_{ji},X_i \right) -T_j \right),
\end{equation}
where $\mathrm{sign}$ is the sign function,   $T_j$ is a threshold associated with the neuron, and the $XNOR$ operation is defined in  Table~\ref{tab:gates}.
Training BNNs is a relatively sophisticated operation,  during which each synapse needs to be associated with a real value in addition to its binary value (see Appendix). Once training is finished, these real values can be discarded, and the neural network is entirely binarized.
Due to their reduced memory requirements, and reliance on simple arithmetic operations, BNNs are especially appropriate for in- or near- memory implementations. 
In particular, multiple groups investigate the implementation of BNN inference with resistive memory tightly integrated at the core of CMOS \cite{bocquet2018,yu2016binary,giacomin2019robust,sun2018xnor,zhou2018new,natsui2018design,tang2017binary, lee2019adaptive}. Usually, resistive memory stores the synaptic weights $W_{ji}$.
However, this comes with a significant  challenge: resistive memory is prone to bit errors, and in digital applications, is typically used with strong error-correcting codes (ECC). 
ECC, which requires large decoding circuits \cite{gregori2003chip}, goes against the principles of in- or near- memory computing.
For this reason,  \cite{bocquet2018} proposes a two-transistor/two-resistor (2T2R) structure, which reduces resistive memory bit errors, without the need for ECC decoding circuit, by storing synaptic weights in a differential fashion.
This architecture allows  the extremely efficient implementation of BNNs, and  using the resistive memory devices in very favorable programming conditions (low energy, high endurance). 
 It should be noted that systems using this architecture function with row-by-row read operations, and do not use the in-memory computing technique of using the Kirchhoff current law to perform the sum operation  of neural networks, while reading all devices at the same time \cite{prezioso2015training,ambrogio2018equivalent}. This choice limits the parallelism of such architectures, while at the same time avoiding the need of analog-to-digital conversion and analog circuits such as operational amplifiers, as discussed in detail in \cite{hirtzlin2019digital}.

In this work, we show that the same architecture can be used for a generalization of BNNs -- ternary neural networks (TNNs)\footnote{
In the literature, the name  ``Ternary Neural Networks'' is sometimes  also used to refer to neural networks where the synaptic weights are ternarized, but the neuronal activations remain real or integer \cite{mellempudi2017ternary,nurvitadhi2017can}.}, 
where neuronal activations and synaptic weights
$A_j$, $X_i$, and $W_{ji}$ can now assume three  values: $+1$, $-1$, and $0$.
Equation~\eqref{eq:activ_BNN} now becomes:
\begin{equation}
\label{eq:activ_TNN}
    A_j = \phi \left( \sum_i  GXNOR \left( W_{ji},X_i \right) -T_j \right).
\end{equation}
$GXNOR$ is the ``gated'' XNOR operation that realizes the product between numbers with values $+1$, $-1$ and $0$ (Table~\ref{tab:gates}).
$\phi$ is an activation function that outputs $+1$ if its input is greater than a threshold $\Delta$, $-1$ if the input is lesser than $-\Delta$ and $0$ otherwise.
We show experimentally and by circuit simulation in sec.~\ref{sec:circuit} how the 2T2R BNN architecture can be extended to TNNs with practically no overhead, in sec.~\ref{sec:programmability} its bit errors, and in sec.~\ref{sec:network} the corresponding benefits in terms of neural network accuracy.

\begin{table}[tbp]
\caption{Truth Tables of the XNOR and GXNOR Gates}
\begin{center}
\begin{tabular}{|c|c|c|}
\hline
$W_{ji}$ & $X_i$ & $XNOR$  \\
\hline
$-1$ & $-1$ & $1$  \\
$-1$ & $1$ & $-1$  \\
$1$ & $-1$ & $-1$  \\
$1$ & $1$ & $1$  \\
\hline
\end{tabular}
\begin{tabular}{|c|c|c|}
\hline
$W_{ji}$ & $X_i$ & $GXNOR$  \\
\hline
$-1$ & $-1$ & $1$  \\
$-1$ & $1$ & $-1$  \\
$1$ & $-1$ & $-1$  \\
$1$ & $1$ & $1$  \\
$0$ & $X$ & $0$  \\
$X$ & $0$ & $0$  \\
\hline
\end{tabular}
\label{tab:gates}
\end{center}
\end{table}


\section{The Operation of A Precharge Sense Amplifier Can Provide Ternary Weights}
\label{sec:circuit}

\begin{figure}[htbp]
	\centering
	\includegraphics[width=3.4in]{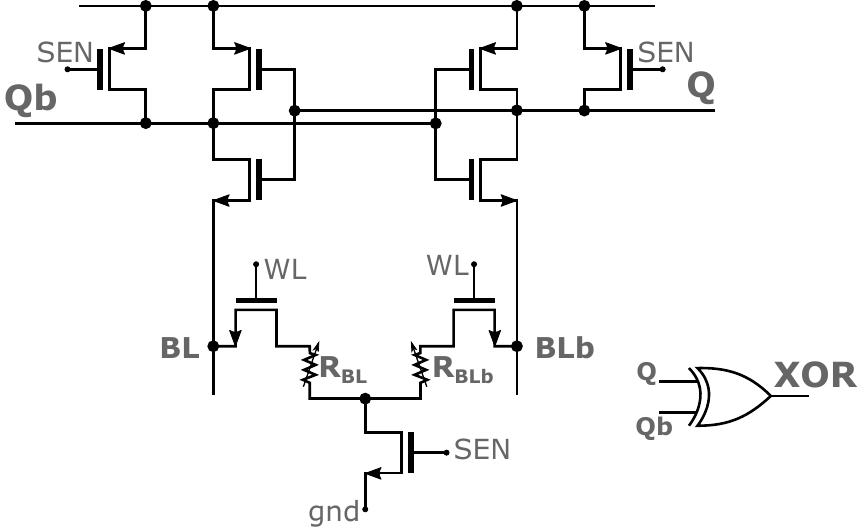}
	\caption{Schematic of the precharge sense amplifier fabricated in the test chip.}
	\label{fig:PCSA}
	\vspace{0.5cm}
\end{figure}

\begin{figure}[htbp]
	\centering
	\includegraphics[width=\linewidth]{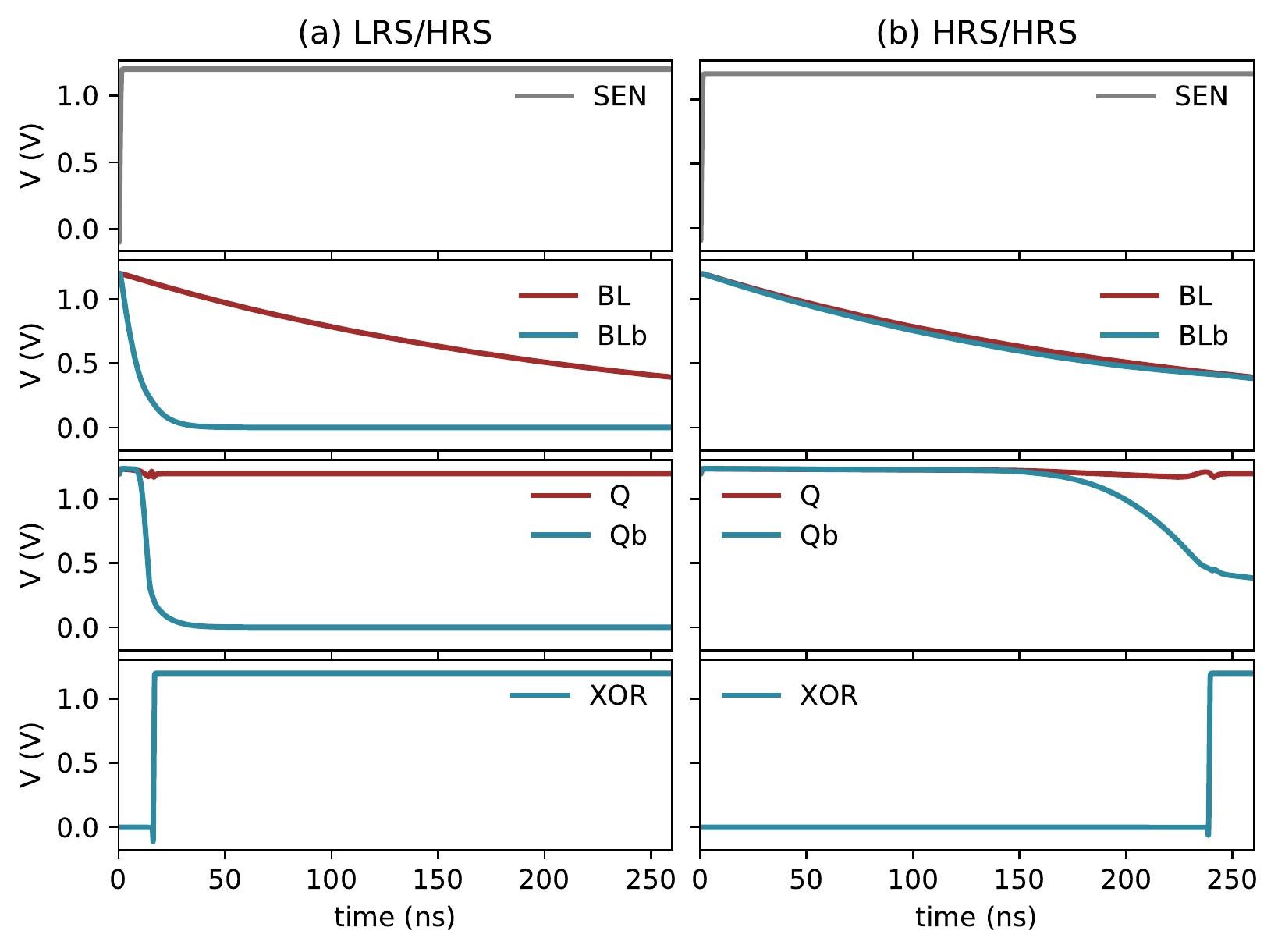}
	\caption{Circuit simulation of the precharge sense amplifier of Fig.~\ref{fig:PCSA} with a supply voltage of $1.2\volt$, using thick oxide transistors (nominal voltage of $5 \volt$)}, if the two devices are programmed in an (a) LRS / HRS ({$5\kilo\ohm$/$350\kilo\ohm$)} or (b) HRS/HRS ({$320\kilo\ohm$/$350\kilo\ohm$}) configuration.
	
	\label{fig:SPICE}
\end{figure}
In this work, we use the architecture of \cite{bocquet2018}, where synaptic weights are stored in a differential fashion.
Each bit is implemented using two devices programmed either as low resistance state (LRS) / high resistance state (HRS) to mean weight $+1$ or HRS/LRS to mean weight $-1$. Fig.~\ref{fig:testchip} presents the test chip used for the experiments. 
This chip cointegrates $130\nano\meter$ CMOS and resistive memory in the back-end-of-line, between levels four and five of metal. The resistive memory cells are based on $10\nano\meter$ thick hafnium oxide (Fig.~\ref{fig:testchip}(a)). 
All devices are integrated with a series NMOS transistor.
After an initial forming step (consisting in the application of a voltage ramp from zero volts to $3.3\volt$ at a rate of $1000\volt\per\second$, and with a current limited to a compliance of $200\micro\ampere$), 
the devices can switch between high resistance state (HRS) and low resistance state (LRS), through the dissolution or creation of conductive filaments of oxygen vacancies. 
Programming into the HRS is obtained by the application of a negative RESET voltage pulse (typically between $1.5\volt$ and $2.5\volt$ during $1\micro\second$). Programming into the LRS is obtained by the application of a positive SET pulse (also typically between $1.5\volt$ and $2.5\volt$ during $1 \micro\second$), with current limited to a compliance current through the choice of the voltage applied on the transistor gate through the word line. 
This test chip is designed with highly conservative sizing, allowing the application of a wide range of voltages and electrical currents to the RRAM cells. The area of each bit cell is $6.6 \times 6.9 \micro\meter ^2$.
More details on the RRAM technology are provided in \cite{hirtzlin2019digital}.

Our experiments are based on a $2,048$ devices array incorporating all sense and periphery circuitry, illustrated in Fig.~\ref{fig:testchip}(b-c).
The ternary synaptic weights are read using  on-chip precharge sense amplifiers (PCSA), presented in Fig.~\ref{fig:PCSA}, and initially proposed  in \cite{zhao2009high} for reading spin-transfer magnetoresistive random access memory.
Fig.~\ref{fig:SPICE}(a) shows an electrical simulation of this circuit to explain its working principle, using the Mentor Graphics Eldo simulator.  
These first simulations are presented in the commercial $130\nano\meter$ ultra-low leakage technology, 
used in our test chip, with a low supply voltage of $1.2 \volt$ \cite{nearth}, with thick oxide transistors (the nominal voltage in this process for thick oxide transistor is $5 \volt$). Since the technology targets ultra-low leakage applications the threshold voltages are significantly high (around $0.6 \volt$), thus a supply voltage of $1.2 \volt$ significantly reduces the overdrive of the transistors ($V_{GS} - V_{TH}$).

In the first phase (SEN=0), the outputs Q and Qb are precharged to the supply voltage $V_{DD}$.
In the second phase (SEN=$V_{DD}$), each branch starts to discharge to the ground. The branch that has the resistive memory (BL or BLb) with the lowest  electrical resistance discharges faster and causes its associated inverter to drive the output of the other inverter to the supply voltage. At the end of the process, the two outputs are therefore complementary and can be used to tell which resistive memory has the highest resistance and therefore the synaptic weight.
We observed that the convergence speed of a PCSA depends heavily on the resistance state of the two resistive memories. 
This effect is particularly magnified when the PCSA is used with a reduced overdrive,  as presented here:
the operation of the sense amplifier is slowed down, with regards to nominal voltage operation, and the convergence speed differences between resistance values become more apparent.
Fig.~\ref{fig:SPICE}(b) shows a simulation where the two devices, BL and BLb, were programmed in the HRS. We see that the two outputs converge to complementary values in more than $200\nano\second$, whereas less than  $50 \nano\second$ were necessary in Fig.~\ref{fig:SPICE}(a), where the devices are programmed in complementary LRS/HRS states. 

These first simulations suggest a technique for implementing ternary weights using the memory array of our test chip. Similarly to when this array is used to implement BNN,  we propose to program the devices in the LRS/HRS configuration to mean the synaptic weight $1$, and HRS/LRS  to mean the synaptic weight $-1$. Additionally, we use the HRS/HRS configuration to mean synaptic weight $0$, while the LRS/LRS configuration is avoided.  The sense operation is performed during a duration of $50\nano\second$. If at the end of this period, outputs Q and Qb have differentiated, causing the output of the XOR gate to be 1, output Q determines the synaptic weight ($1$ or $-1$). Otherwise,  the output of the XOR gate is 0, and the weight is  determined to be $0$. 

This type of coding is reminiscent to the one used by the 2T2R ternary content-addressable memory (TCAM) cell of  \cite{yang2019ternary}, where the LRS/HRS combination is used for coding $0$, the HRS/LRS combination for coding $1$, and the HRS/HRS combination for coding ``don't care'' (or X).

\begin{figure}[tbp]	
\centering	
\includegraphics[width=3.4in]{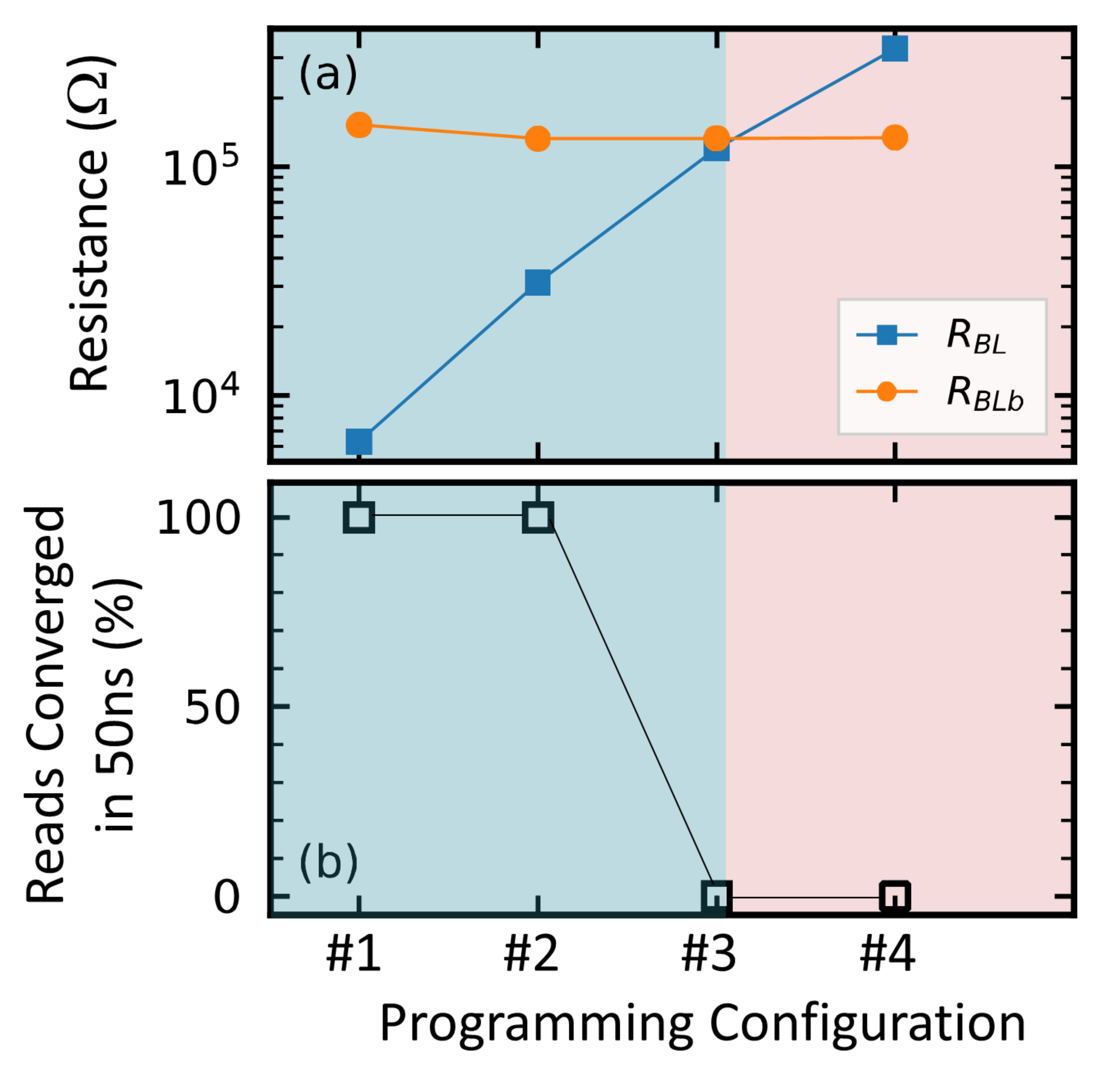} 
\caption{
Two devices have been programmed in four distinct programming conditions, presented in (a), and measured using an on-chip sense amplifier. (b) Proportion of read operations that have converged in $50\nano\second$, over 100 trials.
} 	\label{fig:SenseVsR} \end{figure}

\begin{figure}[tbp]
	\centering 	
	\includegraphics[width=2.8in]{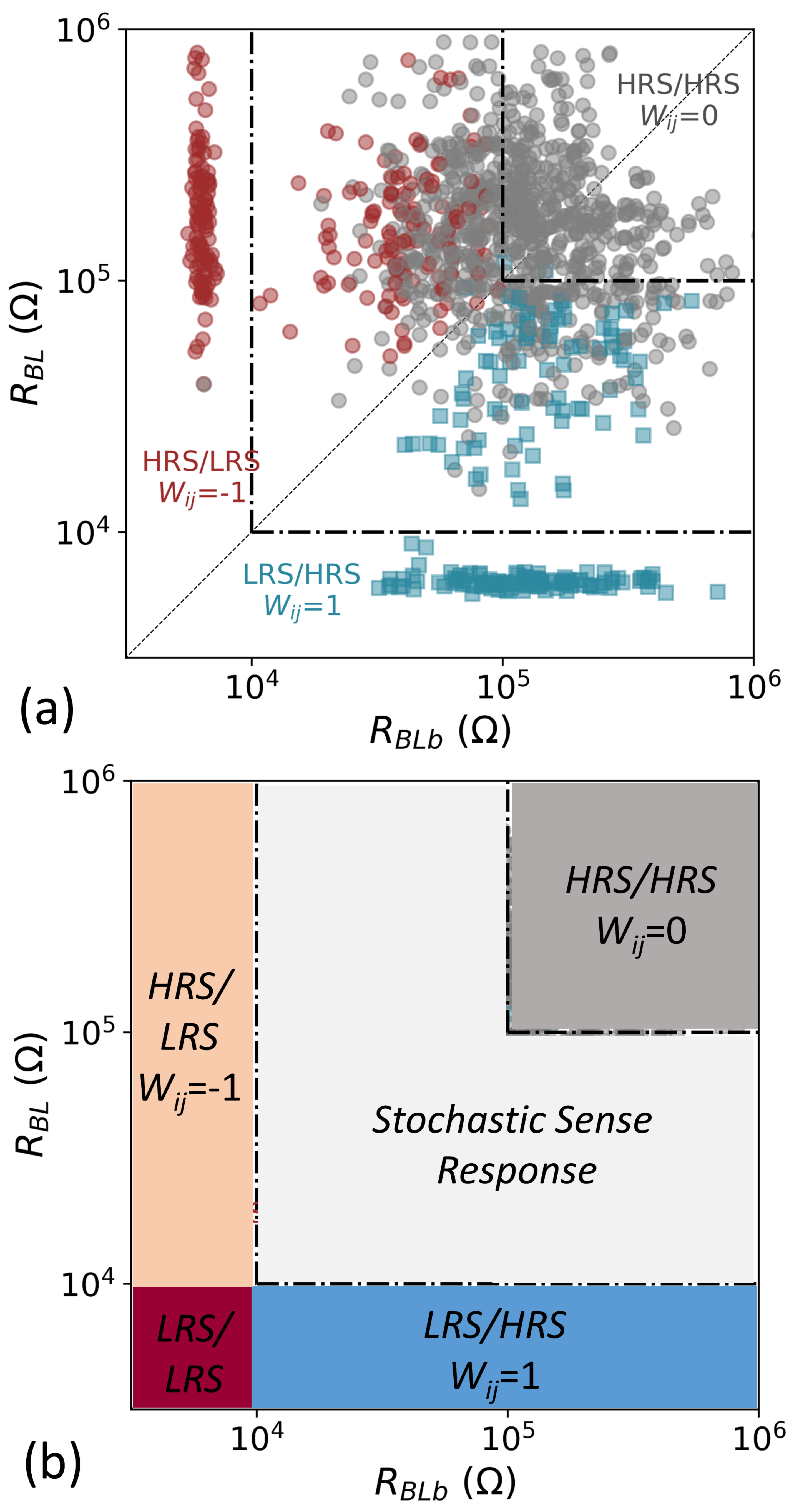} 
	\caption{For 109 device pairs programmed with multiple $R_{BL}/R_{BLb}$ configuration, value of the synaptic weight measured by the on-chip sense amplifier using the  strategy described in  body text and $50\nano\second$ reading time.
	} 	\label{fig:multiniveau_2}
\end{figure}

Experimental measurements on our test chip confirm that the PCSA can be used in this fashion. 
We first focus on one synapse of the memory array. We program one of the two devices (BLb) to a resistance of $100\kilo\ohm$. We then program its complementary device BL to several resistance values, and for each of them perform 100 read operations of duration $50\nano\second$, using on-chip PCSAs.

These PCSAs are fabricated using thick-oxide transistors, designed for a nominal supply voltage of $5V$, and here used with a supply voltage of $1.2V$, close to their threshold voltage ($0.6V$), to reduce their overdrive, and thus to exacerbate the PCSA delay variations.
In the test chip, they are sized conservatively with a total area of $290 \micro\meter ^2$.
The use of thick oxide transistors in this test chip allows us to investigate the behavior of the devices at high voltages, without the concern of damaging the CMOS periphery circuits.
Fig.~\ref{fig:SenseVsR}
plots the probability that the sense amplifier has converged during the read time.
In  $50\nano\second$, the read operation is only converged if the resistance of the BL device is significantly lower than $100\kilo\ohm$.

To evaluate this behavior in a broader range of programming conditions, we repeated the experiment on 109 devices and their complementary devices of the memory array programmed, every 14 times, with various resistance values in the resistive memory, and performed a read operation  in $50\nano\second$ with an on-chip PCSA.
The memory array of our test chip features one separate PCSA per column. Therefore, 32 different PCSAs are used in our results.
Fig.~\ref{fig:multiniveau_2}(a) shows, for each couple of resistance values $R_{BL}$ and $R_{BLb}$ if the read operation was converged with $Q=V_{DD}$ (blue), meaning a weight of $1$, converged with $Q=0$ (red), meaning a weight of $-1$, or not converged (grey) meaning a weight of $0$.

The results confirm that  LRS/HRS or HRS/LRS configurations may be used to mean weights $1$ and $-1$, and HRS/HRS for weight $0$. 
When both devices are in HRS (resistance higher than $100\kilo\ohm$, the PCSA never converges within $50\nano\second$ (weight of $0$). When one device is in LRS (resistance lower than $10\kilo\ohm$, the PCSA always converges within $50\nano\second$ (weight of $\pm1$).
The separation between the $1$ (or $-1$) and $0$ regions is not strict, and for intermediate resistance values, we see that the read operation may or may not converge in $50\nano\second$.  
Fig.~\ref{fig:multiniveau_2}(b) summarizes the different operation regimes of the PCSA.

\begin{table}[h]
\begin{center}
	\vspace{0.5cm}
\caption{Error Rates on Ternary Weights Measured Experimentally}
\begin{tabular}{|l|c|c|c|}
\hline
 Programming & Type 1  & Type 2 & Type 3  
 \\ Conditions& ($1\longleftrightarrow -1$)& ($\pm 1\rightarrow0$) &   ($0 \rightarrow \pm 1$) 
  \\ \hline
Fig.~\ref{fig:distrib}(a) & $<10^{-6}$ &  $<1\%$ &  $6.5\%$
\\Fig.~\ref{fig:distrib}(b) & $<10^{-6}$ &  $<1\%$ &  $18.5\%$
 \\ \hline
\end{tabular}
	\vspace{0.5cm}
\label{table:exp_errors}
\end{center}
\end{table}

\section{
Impact of Process, Voltage, and Temperature Variations}
\label{sec:PVT}

\begin{figure*}[tbp]
	\centering 	
	\includegraphics[width=1.0\textwidth]{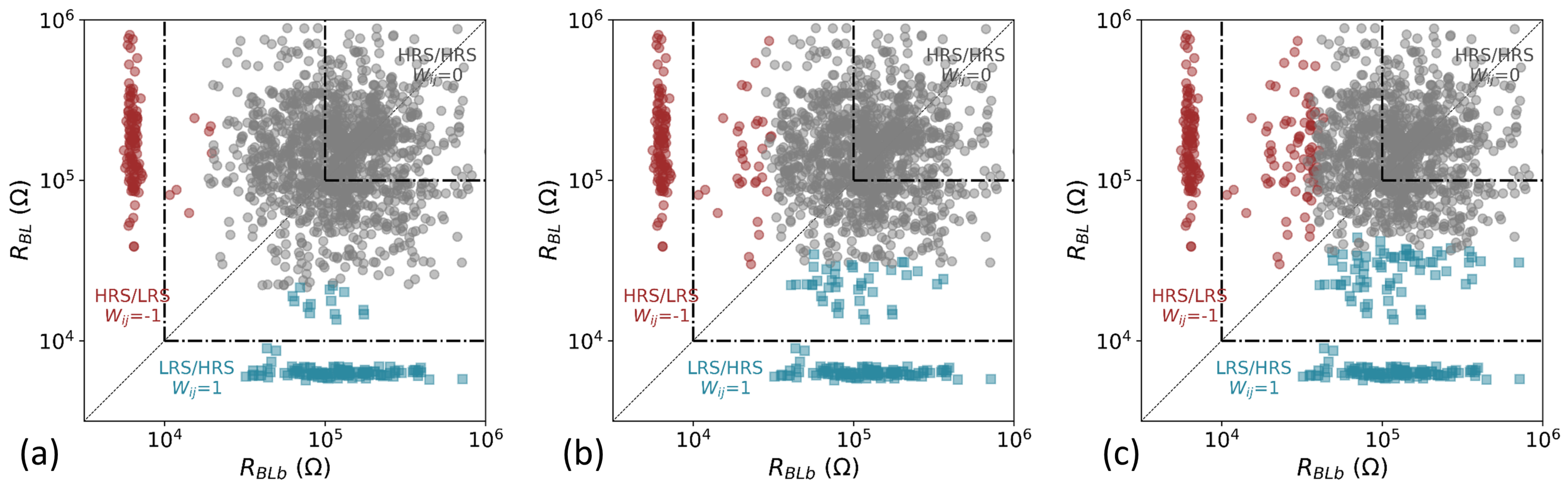} 
    \caption{
    Three Monte Carlo SPICE-based simulation of the experiments of Fig.~\ref{fig:multiniveau_2}, in three situations: (a) slow transistors ($0 \celsius$ temperature,  $1.1\volt$ supply voltage), (b) experimental conditions  ($27 \celsius$ temperature,  $1.2\volt$ supply voltage), (c) fast transistors ($60 \celsius$ temperature,  $1.3\volt$ supply voltage). The simulations include local and global process variations, as well as transistor mismatch, in a way that each point in the Figure is obtained using different transistor parameters. All results are plotted in the same manner and with the same conventions as Fig.~\ref{fig:multiniveau_2}.
	} 	\label{fig:PVT}
\end{figure*}

We now verify the robustness of the proposed scheme to process, voltage, and temperature variation.
For this purpose, we performed extensive circuit simulations of the operation of the sense amplifier, reproducing the conditions of the experiments of Fig.~\ref{fig:multiniveau_2}, using the same resistance values for the RRAM devices, and including process, voltage, and temperature variations. The results of the simulations are processed and plotted using the same format as the experimental results of Fig.~\ref{fig:multiniveau_2}, to ease comparison.

These simulations are obtained using the Monte Carlo simulator provided by the Mentor Graphics Eldo tool with parameters validated on silicon, provided by the design kit of our commercial CMOS process. Each point in the graphs of Fig.~\ref{fig:PVT} therefore features different transistor parameters. We included global and local process variations, as well as transistor mismatch, in order to capture the whole range of transistor variabilities observed in silicon.
In order to assess the impact of voltage and temperature variations, these simulations are presented in three conditions: 
slow transistors ($0 \celsius$ temperature, and $1.1\volt$ supply voltage, Fig.~\ref{fig:PVT}(a)), experimental conditions ($27 \celsius$ temperature, and $1.2\volt$ supply voltage, Fig.~\ref{fig:PVT}(b)), and fast transistors ($60 \celsius$ temperature, and $1.3\volt$ supply voltage, Fig.~\ref{fig:PVT}(c)). 
The RRAM devices are modeled by resistors. Their process variations are naturally included through the use of different resistance values in Fig.~\ref{fig:PVT}. The impact of voltage variation on RRAM is naturally included through Ohm's law, and the impact of temperature variation, which is smaller than on transistors, is neglected.

In all three conditions, the simulation results appear very similar to the experiments. Three clear regions are observed: non-convergence of the sense amplifier within $50ns$ for devices in HRS/HRS, and convergence within this time to a $+1$ or $-1$ value for devices in LRS/HRS and HRS/LRS, respectively. However, the frontier between these regimes is much sharper in the simulations than in the experiments. As the different data points in Fig.~\ref{fig:PVT} differ by process and mismatch variations, this suggests that process variation does not cause the stochasticity observed in the experiments of Fig.~\ref{fig:multiniveau_2}, and that they have little impact in our scheme. 

We also see that the frontier between the different sense regimes in all three operating conditions remains firmly within the $10-100 k\Omega$ range, suggesting that even high variations of voltage ($\pm 0.1V$) and temperature ($\pm 30 ^\circ C$) do not endanger the functionality of our scheme. Logically, in the case of fast transistors, the frontier is shifted toward higher resistances, whereas in the case of slow transistors, it is shifted toward lower resistances. Independent simulations allowed  verifying that this change is mostly due to the voltage variations: the temperature variations have an almost negligible impact on the proposed scheme.

We also observed that the impact of voltage variations increased importantly when reducing the supply voltage. For example, with a supply voltage of $0.7V$ instead of the $1.2V$ value considered here, variations of the supply voltage of $\pm0.1V$ can impact the mean switching delay of the PCSA, by a factor two. The thick oxide transistors used in this work have a nominal voltage of $5V$, and a typical threshold voltage of approximately $0.6V$. Therefore, although our scheme is especially appropriate for supply voltages far below the nominal voltage, it is not necessarily appropriate for voltages in the subthreshold regime, or very close to the threshold voltage.

\section{Programmability of Ternary Weights}
\label{sec:programmability}

\begin{figure}[h]
	\centering
	\includegraphics[width=3.4in]{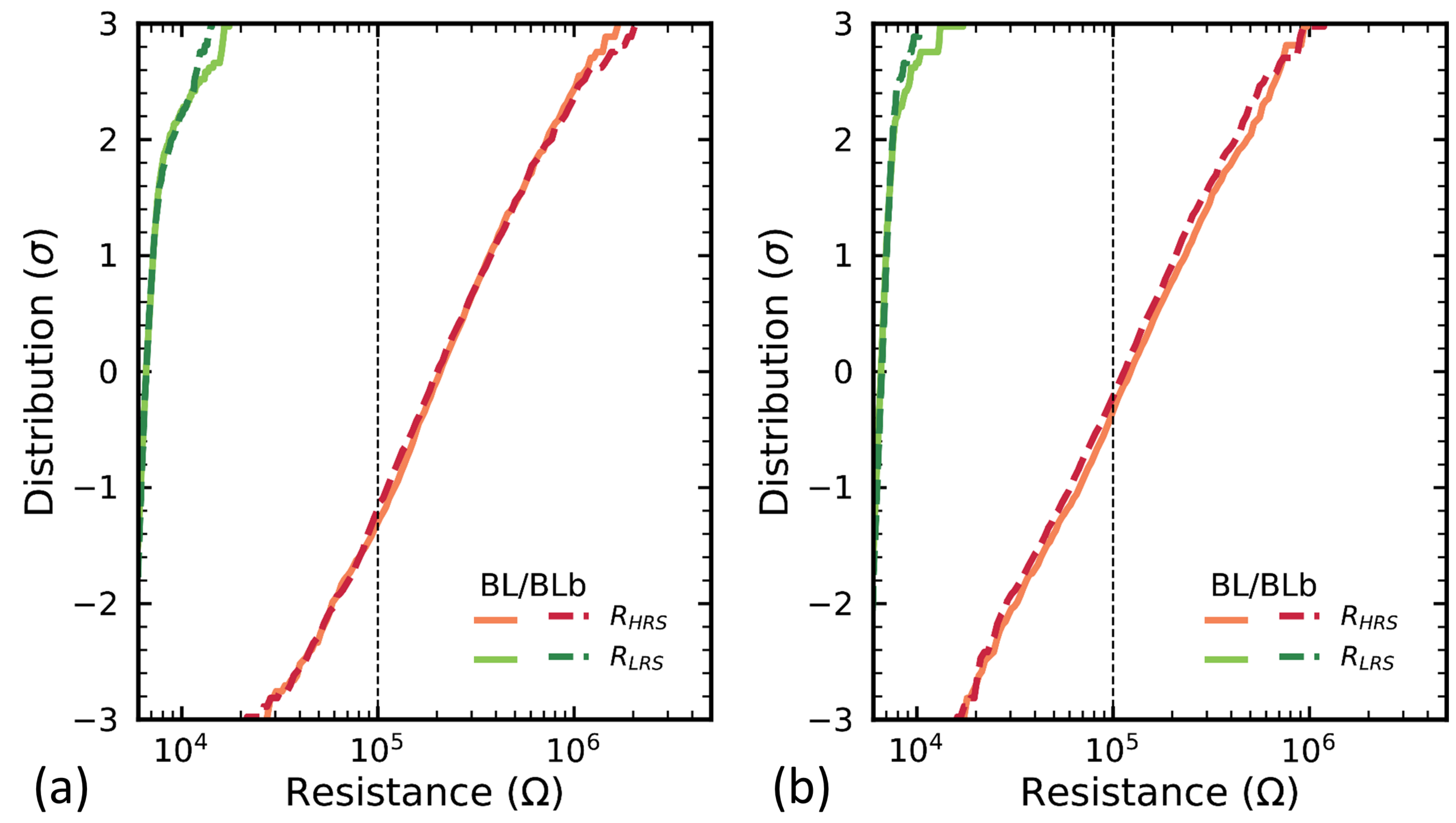}
	\caption{Distribution of the LRS and HRS states programmed with a SET compliance of $200\micro\ampere$, RESET voltage of $2.5\volt$ and programming pulses of (a) $100\micro\second$ and (b) $1\micro\second$. Measurements are performed $2,048$ RRAM devices, separating bit line (full lines) and bit line bar (dashed lines) devices.}
	\label{fig:distrib}
\end{figure}

To ensure reliable functioning of the ternary sense operation,
we have seen that devices in LRS should be programmed to electrical resistance below $10\kilo\ohm$, and devices in HRS to resistances above $100\kilo\ohm$ (Fig.~\ref{fig:multiniveau_2}(b)). The electrical resistance of resistive memory devices depends considerably on their programming conditions \cite{hirtzlin2019digital,grossi2016fundamental}.
Fig.~\ref{fig:distrib} shows the distributions of LRS and HRS resistances using two programming conditions, over the $2,048$ devices of the array, differentiating devices connected to bit lines and to bit lines bar.
We see that in all cases, the LRS features a tight distribution. The SET process is indeed controlled by a compliance current that naturally stops the filament growth at a targeted resistance  value \cite{bocquet2014robust}. An appropriate choice of the compliance current can ensure LRS below $10\kilo\ohm$ in most situations. 

On the other hand, the HRS shows a broad statistical distribution. In the RESET process, the filament indeed breaks in a random process, making it extremely hard to control the final state \cite{bocquet2014robust,ly2018role}. The use of stronger programming conditions leads to higher values of the HRS.  

This asymmetry between the variability of LRS and HRS means that in our scheme, the different ternary weight values feature different error rates naturally.
The ternary error rates in the two programming conditions of Fig.~\ref{fig:distrib}(a) are listed in Table~\ref{table:exp_errors}.
Errors of Type~1,  where weight values of $1$ and $-1$ are inverted are the least frequent.
Errors of Type~2, where a weight value of $1$ or $-1$ is replaced by a weight value of $0$ are infrequent as well.
On the other hand, due to the large variability of the HRS, weight values $0$ have a significant probability to be measured as $1$ or $-1$ (Type~3 errors):  $6.5\%$ in the conditions of Fig.~\ref{fig:distrib}(a), and $18.5\%$ in the conditions of Fig.~\ref{fig:distrib}(b).

Some resistive memory technologies with large memory windows, such as specifically optimized conductive bridge memories \cite{vianello2014resistive}, would feature lower Type~3 error rates. 
Similarly, program-and-verify strategies \cite{lee2012multi,alibart2012high,xu2013understanding} may reduce this error rate. 
Nevertheless, the higher error rate for zeros than for $1$ and $-1$ weights is an inherent feature of our architecture. 
Therefore, in the next section, we  assess the impact of these errors on the accuracy of neural networks.


\begin{figure}[bp]
	\centering
	\includegraphics[width=3.4in]{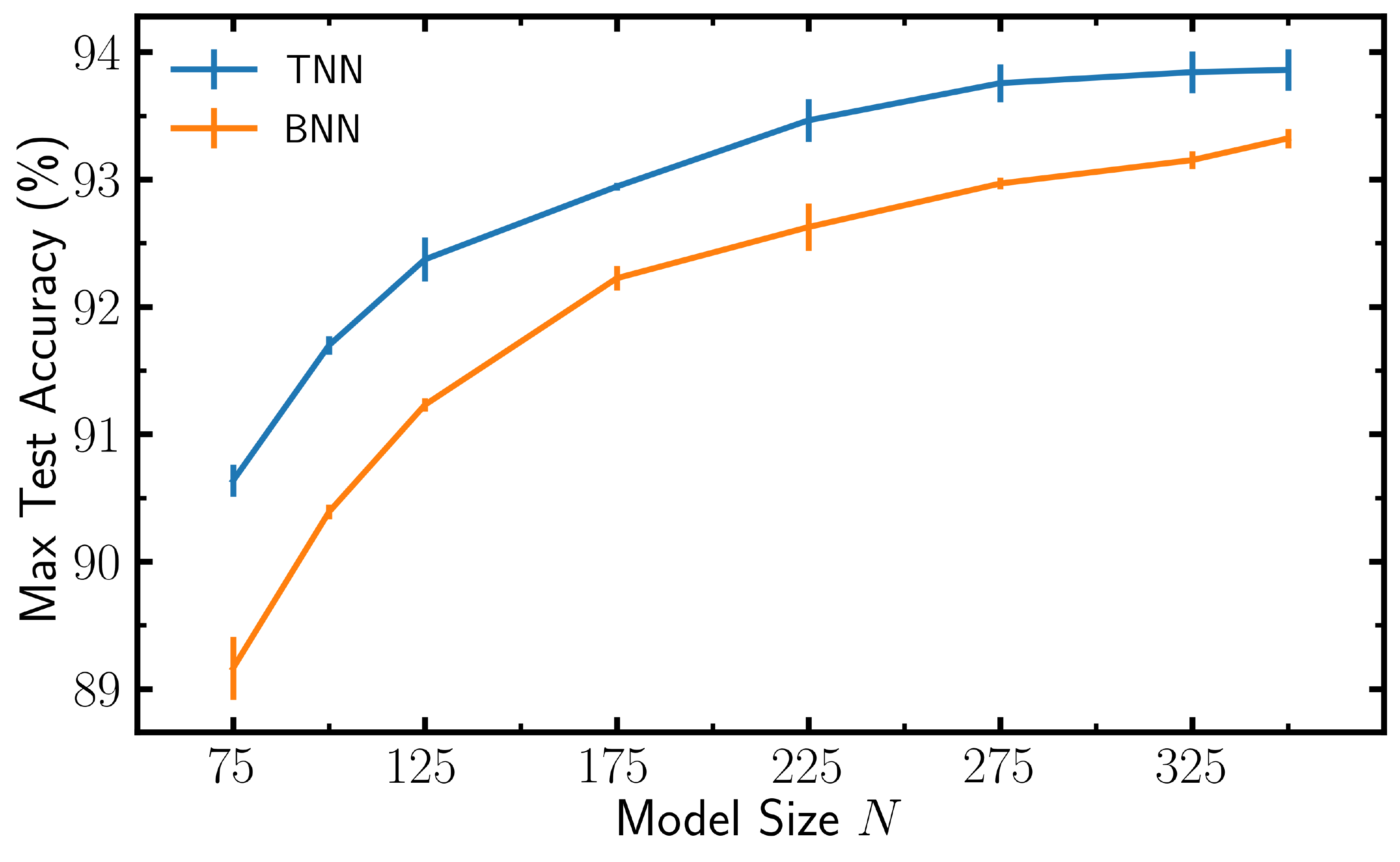}
	\caption{Simulation of the
	maximum test accuracy reached during one training procedure, averaged over five trials, for BNNs and TNNs with various model sizes on the CIFAR-10 dataset. Error bar is one standard deviation.}
	\label{fig:CIFAR10}
\end{figure}

\section{Network-Level Implications}
\label{sec:network}

We first investigate the accuracy gain when using ternarized instead of binarized networks.
We trained BNN and TNN versions of networks with Visual Geometry Group (VGG) type architectures  
\cite{simonyan2014very}
on the CIFAR-10 task of image recognition, consisting in classifying 1,024 pixels color images among ten classes (airplane, automobile, bird, cat, deer, dog, frog, horse, ship, and truck) \cite{krizhevsky2009learning}.
Simulations are performed using PyTorch 1.1.0 \cite{paszke2017automatic} on  a cluster of eight  Nvidia GeForce RTX 2080 GPUs.

The architecture of our networks consists of six convolutional layers with kernel size three. The number of filters  at the first layer is called $N$ and is multiplied by two every two layers. Maximum-value pooling with kernel size two is used every two layers and batch-normalization \cite{ioffe2015batch} every layer. 
The classifier consists of one hidden layer of 512 units. For the TNN, the activation function has a threshold $\Delta = 5\cdot10^{-2}$ (as defined in section~\ref{sec:background}).
The training methods for both the BNN and the TNN are described in the Appendix.
The training is performed using the AdamW optimizer \cite{kingma2014adam, loshchilov2017fixing}, with minibatch size $128$.
The initial learning rate is set to $0.01$, and the learning rate schedule from  \cite{loshchilov2017fixing, loshchilov2016sgdr} (Cosine annealing with two restarts, for respectively $100$, $200$, $400$ epochs) is used, resulting in a total of $700$ epochs. 
Training data is augmented using random horizontal flip, and random choice between cropping after padding and random small rotations.

No error is added during the training procedure, as our device is meant to be used for inference. The synaptic weights encoded by device pairs would be set after the model has been trained on a computer. 

Fig.~\ref{fig:CIFAR10} shows the maximum test accuracy resulting from these training simulations, for different sizes of the model. The error bars represent one standard deviation of the training accuracies. 
TNNs always outperform BNNs with the same model size (and, therefore, the same number of synapses). 
The most substantial difference is seen for smaller model size, but a significant gap remains even for large models. 
Besides, the difference in the number of parameters required to reach a given accuracy for TNNs and BNNs increases with higher accuracies. There is, therefore, a definite advantage to use TNNs instead of BNNs. 

\begin{figure}[tbp]
	\centering
	\includegraphics[width=3.4in]{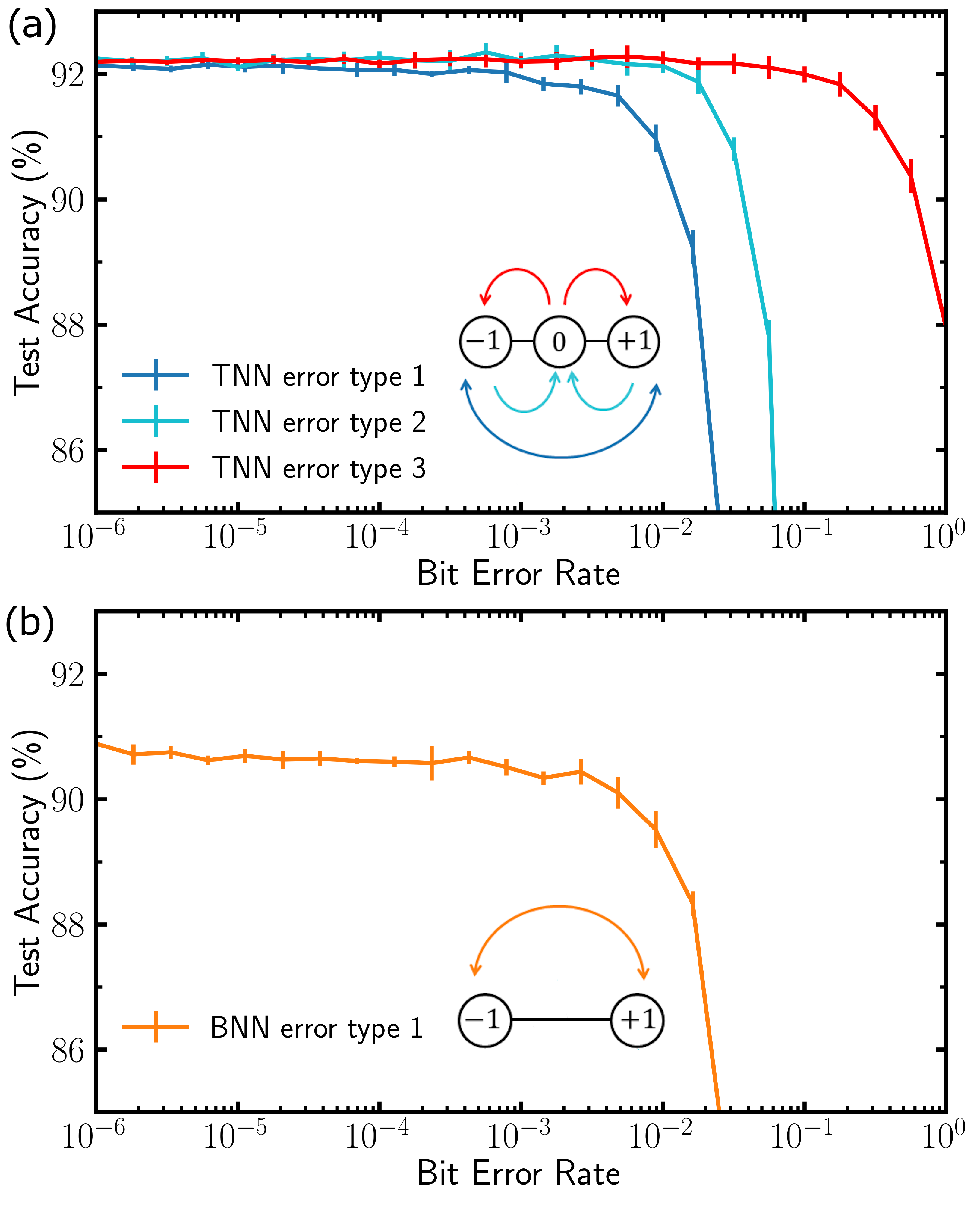}
	\caption{Simulation of the
	impact of Bit Error Rate on the test accuracy at inference time for model size $N = 128$ TNN in (a) and BNN in (b). Type 1 errors are sign switches (e.g. $+1$ mistaken for $-1$), Type 2 errors are
	$\pm 1$ mistaken for $0$, and Type 3 errors are $0$ mistaken for $\pm 1$, as described in the inset schematics. Errors are sampled at each mini batch and the test accuracy is averaged over five passes through the test set. Error bars are one standard deviation. The bit error rate is given as an absolute rate.}
	\label{fig:BT_errors}
\end{figure}

Fig.~\ref{fig:CIFAR10} compared fully ternarized (weights and activations) with regards to fully binarized (weights and activations) ones. Table~\ref{table:acc_gain} lists the impact of weight ternarization for different types of activations (binary, ternary, and real activation).
All results are reported  on a model of size $N=128$, trained  on CIFAR-10, and are averaged over five training procedures.
We observe that for BNNs and TNNs with quantized activations, the accuracy gains provided by ternary weights over binary weights are $0.84$ and $0.86$ points and are statistically significant over the standard deviations. This accuracy gain is more important than the  gain provided by ternary activations over binary activations, which is about $0.3$ points. 
This bigger impact of weight ternarization over ternary activation may come from the ternary kernels having a better expressing power over binary kernels, which are often redundant in practical settings \cite{courbariaux2016binarized}.
The gain of ternary weights drops to $0.26$ points if real activation is allowed (using rectified linear unit, or ReLU, as activation function, see appendix), and is not statistically significant considering the standard deviations. 

Quantized activations are vastly more favorable in the context of hardware implementations, and in this situation, there is thus a statistically significant benefit provided by ternary weights over binary weights.

\begin{table}[!h]
\begin{center}
\vspace{0.5cm}
\caption{Comparison of the gain in test accuracy for a $N=128$ model size on CIFAR-10 obtained by weight ternarization instead of binarization for three types of activation quantization. }
\begin{tabular}{|l|c|c|c|}
\hline
& \multicolumn{3}{c|}{\textbf{Activations}}   \\
 & Binary & Ternary & Full Precision \\ \hline
 \textbf{Weights} & & & \\
Binary                             & $91.19\pm0.08$   & $91.51 \pm 0.09$    & $93.87 \pm 0.19$           \\
Ternary                            & $92.03\pm 0.12$   & $92.35 \pm 0.05$    & $94.13 \pm 0.10$           \\
\textit{Gain of ternarization}              
& \textit{0.84 }   
& \textit{0.86}     
& \textit{0.26}            \\ \hline
\end{tabular}
\vspace{0.5cm}
\label{table:acc_gain}
\end{center}
\end{table}

We finally investigate the impact of bit errors in BNNs and TNNs to see if the advantage provided by using TNNs in our approach remains constant when errors are taken into account. Consistently with the results reported in section~\ref{sec:programmability}, three types of errors are investigated: Type 1 errors are sign switches,
e.g., $+1$ mistaken for $-1$, 
Type 2 errors are only defined for TNNs and correspond to  $\pm1$ mistaken for $0$, and Type 3 errors are $0$ mistaken for $\pm 1$, as illustrated in the inset schematic of Fig.~\ref{fig:BT_errors}(a). 

Fig.~\ref{fig:BT_errors}(a) shows the impact of these  errors on the test accuracy for different values of the  error rate at inference time. These simulation results are presented on CIFAR-10    with a model size of $N=128$. Errors are randomly and artificially introduced in the weights of the neural network .
Bit errors are included at the layer level and sampled at each mini-batch of the test set. Type 1 errors switch the sign of a synaptic weight with a probability equal to the rate of type 1 errors. Type 2 errors set a non-zero synaptic weight to 0 with a probability equal to the type 2 error rate. Type 3 errors set a synaptic weight of 0 to $\pm1$ with a probability equal to the type 3 error rate, the choice of the sign ($+1$ or $-1$) is made with 0.5 probability. Fig.~\ref{fig:BT_errors} is obtained by averaging the test accuracy obtained for five passes through the test set for increasing bit error rate.

Type~1 errors have the most impact on neural network accuracy. As seen in Fig.~\ref{fig:BT_errors}(b), the impact of these errors is similar to the impact of weight errors in a BNN.
On the other hand, Type~3 errors have the least impact, with bit error rates as high as $20\%$ degrading surprisingly little the accuracy.
This result is fortunate, as we have seen in section~\ref{sec:programmability} that Type~3 errors are the most frequent in our architecture. 

We also performed simulations considering all three types of error at the same time, with error rates reported in Table~\ref{table:exp_errors} corresponding to the programming conditions of Fig.~\ref{fig:distrib}(a) and \ref{fig:distrib}(b).  
For Type~1 and Type~2 errors, we considered the upper limits listed in Table~\ref{table:exp_errors}.  
For the conditions of Fig.~\ref{fig:distrib}(a) (Type~3 error rate of 6.5\%), the test accuracy was degraded from 92.2\% to $92.05 \pm 0.14 \%$, and to $92.02 \pm 0.17 \%$ for the conditions of \ref{fig:distrib}(b) (Type~3 error rate of 18.5\%), where the average and standard deviation is performed over 100 passes through the test set. 
We found that the slight degradation on CIFAR-10 test accuracy was mostly due to the Type~2 errors, although Type 3 errors are much more frequent. 

The fact that mistaking a $0$ weight for a $\pm1$ weight (Type~3 error) has much less impact than mistaking a $\pm1$ weight for a $0$ weight (Type~2 error) can seem surprising.
However, it is known, theoretically and practically, that in BNNs, some weights have  little importance to the accuracy of the neural networks \cite{laborieux2020synaptic}. They typically correspond to synapses that feature a $0$ weight in a TNN, whereas synapses with $\pm1$ weights in a TNN correspond to ``important'' synapses of a BNN. It is thus understandable that errors on such synapses have more impact on the final accuracy of the neural network.


\section{
Comparison with Three-Level Programming}
\label{sec:three-level}

An alternative approach to implementing ternary weights with resistive memory can be to program the individual devices into there separate levels. This idea is feasible, as the resistance level of the LRS can to a large extent be controlled through the choice of the compliance current during the SET operation in many resistive memory technologies \cite{bocquet2014robust,hirtzlin2019digital}.  

The obvious advantage of this approach is that it requires a single device per synapse. This idea also brings several challenges. First, the sense operation has to be more complex. The most natural technique is to perform two sense operations, comparing the resistance of a device under test to  two different thresholds. Second, this technique is much more prone to bit errors than our technique, as states are not programmed in a differential fashion \cite{hirtzlin2019digital}. Additionally, this approach does not feature the natural resilience to Type~1 and Type~2 errors, and Type~2 and Type~3 errors will typically feature similar rates. Finally, unlike ours, this approach is prone to resistive drift, inherent to some resistive memory technologies \cite{li2012resistance}.

These comments suggest that the choice of a technique for storing ternary weights should be dictated by technology. Our technique is especially appropriate for resistive memories not supporting single-device multilevel storage, with high error rates, or resistance drift. The three-levels per devices approach would be the most appropriate with devices with well controlled analog storage properties.


\section{Conclusion}

In this work, we revisited a differential memory architecture for BNNs.
We showed experimentally on a hybrid CMOS/RRAM chip that,
its sense amplifier can differentiate not only the LRS/HRS and HRS/LRS states, but also the HRS/HRS states in a single sense operation. 
This feature allows the architecture to store ternary weights, and to provide a building block for ternary neural networks. 
We showed by neural network simulation on the CIFAR-10 task the benefits of using ternary instead of binary networks, and the high resilience of TNNs to weights errors, as the type of errors observed experimentally in our scheme is also the type of errors to which TNNs are the most immune. 
This resilience allows the use of our architecture without relying on any formal error correction.
Our approach also appears resilient to process, voltage, and temperature variation if the supply voltage remains reasonably higher than the threshold voltage of the transistors.

As this behavior of the sense amplifier is exacerbated at supply voltages below the nominal voltage, our approach especially targets extremely energy-conscious applications such as uses within wireless sensors or medical applications. This work opens the way for increasing the edge intelligence in such contexts, and also highlights that the low voltage operation of circuits may sometimes provide opportunities for new functionalities.

\section*{Acknowledgments}
The authors would like to thank M.~Ernoult and L.~Herrera Diez for fruitful discussions.


\section*{Appendix: Training Algorithm of Binarized and Ternary Neural Networks}

\begin{algorithm}[ht!]{\emph{Input}: $W^{\rm h}$, $\theta^{\rm BN} = (\gamma_l, \beta_l)$, $U_W$, $U_{\theta}$, $(X, y)$, $\eta$. \\
\emph{Output}: $W^{\rm h}$, $\theta^{\rm BN}$, $U_W$, $U_{\theta}$.}
\caption{Training procedure for binary and ternary neural networks. $W^{\rm h}$ are the hidden weights, $\theta^{\rm BN} = (\gamma_l, \beta_l)$ are Batch Normalization parameters, $U_W$ and $U_{\theta}$ are the parameter updates prescribed by the Adam algorithm \cite{kingma2014adam}, $(X, y)$ is a batch of labelled training data, and $\eta$ is the learning rate. ``cache'' denotes all the intermediate layers computations needed to be stored for the backward pass. ${\rm Quantize}$ is either $\phi$ or ${\rm sign}$ as defined in section \ref{sec:background}. ``~$\cdot$~'' denotes the element-wise product of two tensors with compatible shapes.}\label{algo}
\begin{algorithmic}[1]
\State $W^{\rm Q} \leftarrow {\rm Quantize} ( W^{\rm h})$ \Comment{Computing quantized weights}
\State $A_0 \leftarrow X$ \Comment{Input is not quantized}
\For{$l=1$ to $L$} \Comment{For loop over the layers}
\State $z_l \leftarrow W^{\rm Q}_l A_l$ \Comment{Matrix multiplication}
\State $A_l \leftarrow \gamma_l \cdot \frac{z_l - {\rm E}(z_l)}{\sqrt{{\rm Var}(z_l) + \epsilon}} + \beta_l$ \Comment{Batch~Normalization~\cite{ioffe2015batch}}
\If{$l < L$} \Comment{If not the last layer}
\State $A_l \leftarrow {\rm Quantize}(A_l)$  \Comment{Activation is quantized}
\EndIf
\EndFor
\State $\hat{y} \leftarrow A_L$
\State $C \leftarrow {\rm Cost}(\hat{y}, y)$ \Comment{Compute mean loss over the batch}
\State $(\partial_W C, \partial_{\theta} C) \leftarrow {\rm Backward}(C, \hat{y}, W^{\rm Q}, \theta^{\rm BN}, {\rm cache}) $ \Comment{Cost gradients}

\State $ (U_W, U_{\theta}) \leftarrow {\rm Adam}(\partial_W C, \partial_{\theta} C, U_W, U_{\theta} )$ 
\State $W^{\rm h} \leftarrow W^{\rm h} - \eta  U_W $
\State $\theta^{\rm BN} \leftarrow \theta^{\rm BN} - \eta U_{\theta}$ \\
\Return $W^{\rm h}$, $\theta^{\rm BN}$, $U_W$, $U_{\theta}$
\end{algorithmic}
\end{algorithm}


During the training of BNNs and TNNs, each quantized (binary or ternary) weight is associated with a real hidden weight. This approach to training quantized  neural network was introduced  in \cite{courbariaux2016binarized} and is presented in Algorithm~\ref{algo}. 

The quantized weights are used for computing neuron values (equations~\eqref{eq:activ_BNN} and~\eqref{eq:activ_TNN}), as well as the gradients values in the backward pass.
However, training steps are achieved  by updating the real hidden weights. The quantized weight is then determined by applying to the real value the quantizing function ${\rm Quantize}$, which is $\phi$ for ternary or ${\rm sign}$ for binary as defined in section \ref{sec:background}. The quantization of activations is done by applying the same function ${\rm Quantize}$, except for real activation, which is done by applying a rectified linear unit (${\rm ReLU}(x) = {\rm max}(0, x)$).

Quantized activation functions ($\phi$ or ${\rm sign}$) have zero derivatives almost everywhere, 
which is an issue for backpropagating the error gradients through the network. 
A way around this issue is the use of a straight-through estimator \cite{bengio2013estimating}, which consists in taking the derivative of another function instead of the almost everywhere zero derivatives. 
Throughout this work, we take the derivative of $\rm Hardtanh$, which is 1 between -1 and 1 and 0 elsewhere, both for binary and ternary activations. 

The simulation code used in this work is available publicly in the Github repository: \url{https://github.com/Laborieux-Axel/Quantized_VGG}

\FloatBarrier


\begin{IEEEbiography}[{\includegraphics[width=1in,height=1.25in,clip,keepaspectratio]{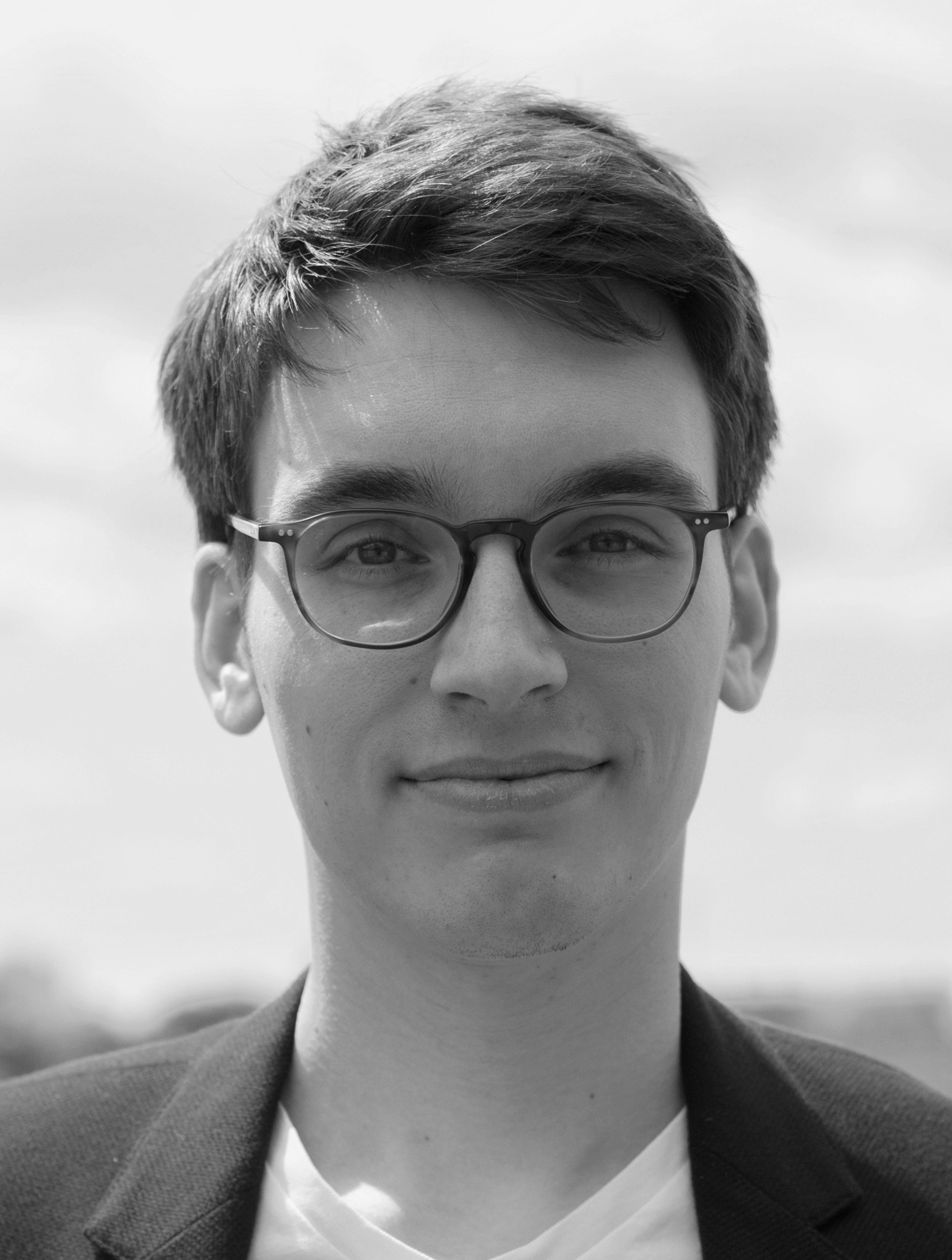}}]{Axel Laborieux}
Axel Laborieux received the M.S. degree in condensed matter physics from the Universit\'e Paris-Saclay, France, in 2018, where he is currently pursuing the Ph.D. degree in neuromorphic computing. His research interest includes the benefits brought by complex synapse behaviors in binarized neural networks and their physical implementation using spintronic nanodevices.
\end{IEEEbiography}


\begin{IEEEbiography}[{\includegraphics[width=1in,height=1.25in,clip,keepaspectratio]{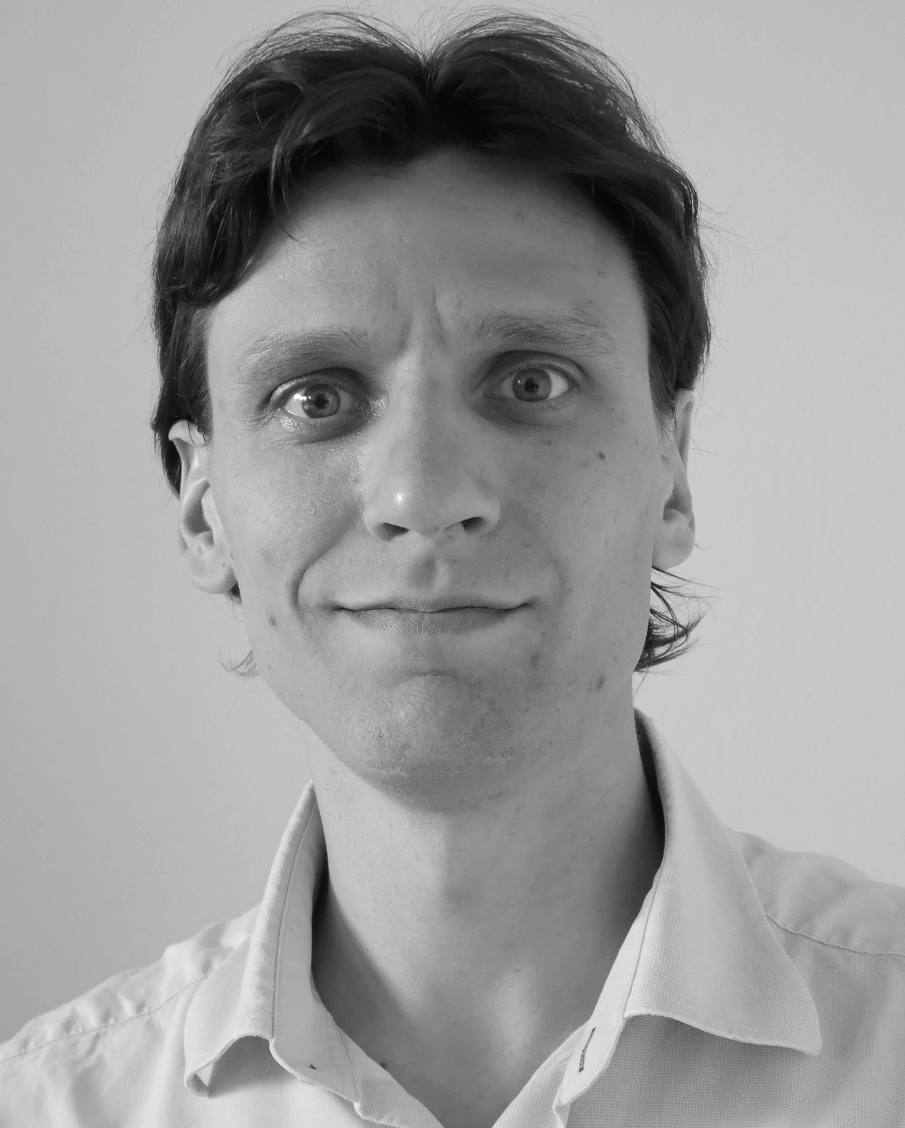}}]{Marc Bocquet}
Marc Bocquet received the M.S. degree in electrical engineering and the Ph.D. degree in electrical engineering from the University of Grenoble, France, in 2006 and 2009, respectively. He is currently an Associate Professor with the Institute of Materials, Microelectronics, and Nanosciences of Provence, IM2NP, Universit\'e of Aix-Marseille and Toulon. His research interests include memory model, memory design, characterization, and reliability.
\end{IEEEbiography}

\begin{IEEEbiography}[{\includegraphics[width=1in,height=1.25in,clip,keepaspectratio]{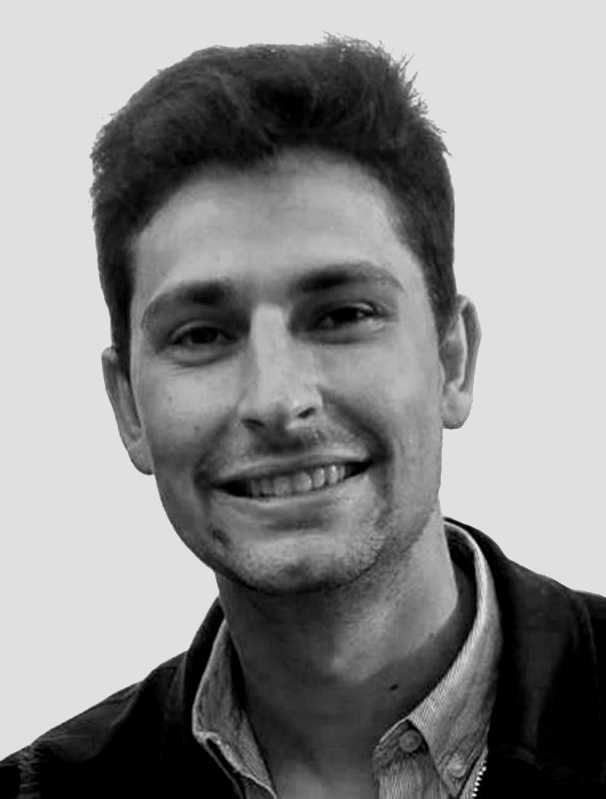}}]{Tifenn Hirtzlin}
Tifenn Hirtzlin received the M.S. degree in nanosciences and electronics from the Universit\'e Paris-Sud, France, in 2017, where he is currently pursuing the Ph.D. degree in electrical engineering. His research interest includes designing intelligent memory-chip for low energy hardware data processing using bio-inspired concepts as a probabilistic approach to brain function and more conventional neural network approaches.
\end{IEEEbiography}

\begin{IEEEbiography}[{\includegraphics[width=1in,height=1.25in,clip,keepaspectratio]{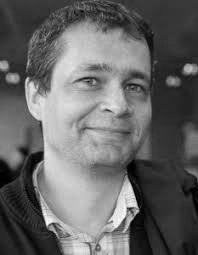}}]{Jacques-Olivier Klein}
Jacques-Olivier Klein (M’90) received the Ph.D. degree from the Universit\'e Paris-Sud, France, in 1995, where he is currently a Full Professor. He focuses on the architecture of circuits and systems based on emerging nanodevices in the field of nanomagnetism and bio-inspired nanoelectronics. He is also a Lecturer with the Institut Universitaire de Technologie (IUT), Cachan. He has authored more than 100 technical papers.
\end{IEEEbiography}

\begin{IEEEbiography}[{\includegraphics[width=1in,height=1.25in,clip,keepaspectratio]{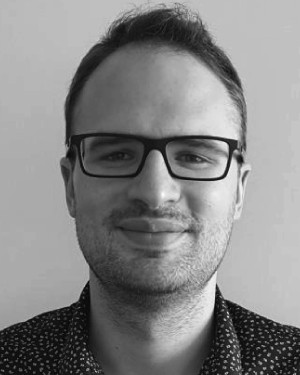}}]{Etienne Nowak}
Etienne Nowak received the M.Sc. degree in microelectronics from Grenoble University, Grenoble, France; Polito di Torino, Turin, Italy; and the Ecole Polytechnique F\'ed\'erale de Lausanne, Lausanne, Switzerland, in 2007, and the Ph.D. degree from the Institut National Polytechnique de Grenoble, Grenoble, France, in 2010.
From 2010 to 2014, he was a Senior Engineer at the Semiconductor Research and Development Center, Samsung Electronics, Hwaseong, South Korea, where he was involved in the first generations of vertical nand flash memory. He joined CEA-Leti, Grenoble, France, in 2014, as a Project Manager on emerging nonvolatile memory. He published over 30 papers and holds two patents on these topics. Since 2017, he has been appointed as the Head of the Advanced Memory Device Laboratory, CEA-Leti, Grenoble, France, dedicated to nonvolatile memory backend technologies.
\end{IEEEbiography}

\begin{IEEEbiography}[{\includegraphics[width=1in,height=1.25in,clip,keepaspectratio]{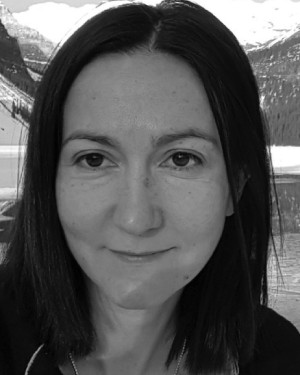}}]{Elisa Vianello}
Elisa Vianello received the Ph.D. degree in
microelectronics from the University of Udine,
Udine, Italy, and the Polytechnic Institute of
Grenoble, Grenoble, France, in 2009.
She has been a Scientist with the Laboratoire d’Electronique des Technologies de
l’Information, Commissariat à l’Energie Atomique et aux Energies Alternatives, Grenoble,
since 2011. Her current research interests
include resistive switching memory devices and
selectors and the use of nanotechnologies for
memory-centric computing and neuromorphic systems.
\end{IEEEbiography}

\begin{IEEEbiography}[{\includegraphics[width=1in,height=1.25in,clip,keepaspectratio]{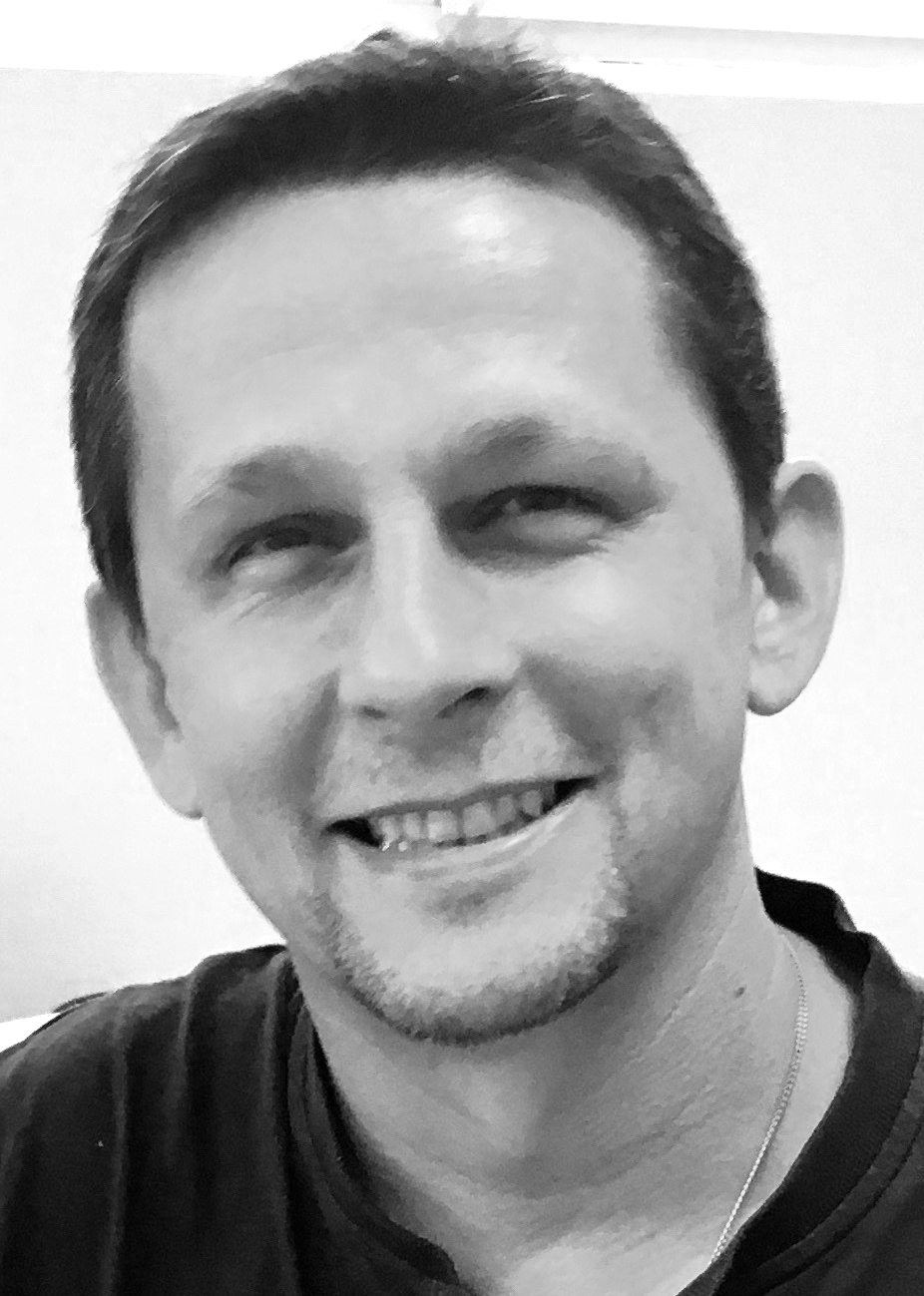}}]{Jean-Michel Portal}
Jean-Michel Portal graduated in Electronic Engineering in 1996 and received the Ph.D. degree in Computer Sciences in 1999.  is currently a Full professor in Electronics at Aix-Marseille University, where he heads the electronic department of the Institute of Materials, Microelectronics, and Nanosciences of Provence (IM2NP). His research interests include emerging non-volatile memory design and neuromorphic applications. He is author or co-author of more than 200 articles in International Refereed Journals and Conferences, and is a co-inventor of six patents. He has supervised 20 Ph.D. students. He is a recipient of the NanoArch 2012, Newcas 2013 and IEEE Transactions on Circuits and Systems Guillemin-Cauer 2017 Best Paper Awards.
\end{IEEEbiography}

\begin{IEEEbiography}[{\includegraphics[width=1in,height=1.25in,clip,keepaspectratio]{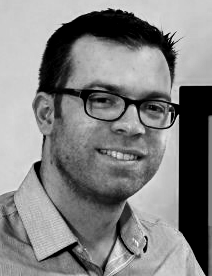}}]{Damien Querlioz}
Damien Querlioz (M’09) received the predoctoral education from the Ecole Normale Sup\'erieure, Paris, and the Ph.D. degree from Universit\'e Paris-Sud, in 2008. After postdoctoral appointments at Stanford University and CEA, he became a Permanent Researcher with the Centre for Nanoscience and Nanotechnology of CNRS and Universit\'e Paris-Saclay. He focuses on novel usages of emerging non-volatile memory, in particular relying on inspirations from biology and machine learning. He coordinates the INTEGNANO Interdisciplinary Research Group.  In 2016, he was a recipient of the European Research Council Starting Grant to develop the concept of natively intelligent memory. In 2017, he received the CNRS Bronze Medal.
\end{IEEEbiography}

\end{document}